\documentclass[12pt]{article}
\usepackage{amsfonts,amsmath,amssymb}
\usepackage{scalerel}[2016/12/29]
\usepackage{graphicx}
\usepackage{tikz}
\usepackage{braket}

\usetikzlibrary{arrows}
\usetikzlibrary{decorations.markings}

\numberwithin{equation}{section}

\newcommand{\e}{\epsilon}
\newcommand{\ro}{\rho(\epsilon)}

\def\be{\begin{equation}}
\def\ee{\end{equation}}
\def\bea{\begin{eqnarray}}
\def\eea{\end{eqnarray}}

\def\half{{\displaystyle {1\over 2}}}

\def\bb{\hskip -0.5mm}

\def\rme{\mathrm{e}}

\usepackage{amsmath}
\usepackage{epsfig,graphicx}
\usepackage{graphicx}
\usepackage{tikz}

\def\be{\begin{equation}}
\def\ee{\end{equation}}
\def\bea{\begin{eqnarray}}
\def\eea{\end{eqnarray}}

\def\rme{\mathrm{e}}

\begin{document}

\thispagestyle{plain}

\title{\bf\large The role of the density of states in Bose-Einstein condensation\\
}

\author{Alexios P. Polychronakos$^\dagger$ and St\'ephane Ouvry$^*$}

\date{ }

\maketitle

{\em
{\centerline{$\dagger\bb$ Physics Department, the City College of New York, NY 10031, USA}}
\vskip 0.1cm
\centerline{\it apolychronakos@ccny.cuny.edu}
\vskip .2 cm \centerline{and}
\vskip .2 cm
{\centerline{The Graduate Center, CUNY, New York, NY 10016, USA}
\vskip .1 cm
\centerline{\it apolychronakos@gc.cuny.edu}
\vskip 0.5cm
{\centerline{* LPTMS, CNRS, Universit\'e Paris-Saclay, 91405 Orsay Cedex, France}}
\vskip .1 cm \centerline{\it stephane.ouvry@universite-paris-saclay.fr}
}}

\maketitle

\begin{abstract}

The onset of Bose-Einstein condensation in systems with { various} densities of states is examined, with particular
attention to the role of the behavior of their {energy} spectrum at low and high energies.
Specifically, the results of Chatterjee and
Diaconis, which rely exclusively on the high-energy behavior, are compared and reconciled with those of the standard physics approach, where the existence of condensation is determined by the low-energy behavior.

\end{abstract}

\vskip 1cm

\vfill
\eject


\section{Introduction}

Bose-Einstein (BE) condensation, predicted 100 years ago and experimentally observed 30 years ago  \cite{Ketterle,Cornell},
remains to this day a fascinating topic, as one of the rare occasions where a system
manifests macroscopic quantum behavior. Its theoretical study, in particular, sharpens our understanding of
quantum statistical mechanics and the relation between ensembles in the thermodynamic limit.

In an interesting paper, Chatterjee and Diaconis (CD)  \cite{Diaconis} considered BE condensation in the canonical
(rather than grand canonical) setting and calculated the fluctuations of the number of particles condensing in the
ground state. A basic result of their analysis is that, as was assumed earlier by H.D. Politzer  \cite{Politzer}, when BE condensation occurs
the ground state acts as a particle reservoir for the remaining
energy levels, which can then be treated in the grand canonical formulation. { (This is a corollary of
Lemma 3.4 in CD and the fact that, when condensation happens, the fluctuations of the number of particles $N_0$ in the
ground state are much smaller than its average and thus the constraint  $M := N - N_0 \ge 0$  in the lemma is
essentially always satisfied.)}

A key element in the CD analysis is the high-energy behavior of the energy spectrum. Expressed in
physics language, their assumption is that the density of states $\rho(\epsilon)$
at energy $\epsilon \to \infty$ behaves as
\be
\rho(\epsilon) \sim L\, \epsilon^{\alpha-1} ~,~~~\alpha\ge 1 \,,\quad \text{as}\quad \epsilon \to \infty
\ee
with $L$ a constant. Given this behavior, and assuming that there is a gap above the ground state, CD's conclusion is that BE condensation will occur. In CD's own words
``no other feature of the energy spectrum is relevant."

The above creates an obvious tension with the standard physics approach. Using the grand canonical ensemble
(assumed to be equivalent to the canonical one for energy levels other than the ground state),
the total number of particles is
\be
N = \int_{\epsilon_0}^\infty {\rho(\epsilon)\, d\epsilon \over e^{\beta(\epsilon - \mu)} -1}
\ee
with $\epsilon_0$ the ground state energy, $\beta = 1/(k_{_B} T)$ the inverse temperature, and
$\mu$ the chemical potential. BE condensation occurs when the grand canonical ensemble cannot accommodate
the actual number of particles { $N$}  and the surplus particles condense in the ground state.
Choosing $\epsilon_0 = 0$, the maximal number of particles occurs for the maximal chemical potential $\mu = \epsilon_0 = 0$.
Thus, if
\be
N_\text{max} = \int_0^\infty  {\rho(\epsilon)\, d\epsilon \over e^{\beta\epsilon} -1} < \infty
\label{Nmax}\ee
there is a maximal particle number that can be accommodated by the ensemble and BE condensation
will occur { for $N > N_\text{max}$}. The contribution to the integral in (\ref{Nmax})
from energies $\epsilon \to \infty$  is always
finite due to the exponential suppression in the denominator (strings, where $\rho(\epsilon)
\sim \exp({\beta_{_{\small H}} \epsilon})$ as $\epsilon \to \infty$, with $\beta_{_H}$ the inverse Hagedorn temperature,
are a notable exception). Thus, finiteness of the integral (\ref{Nmax})
depends solely on the behavior of $\rho(\epsilon)$ near $\epsilon = 0$: if
\be
\rho(\epsilon) \sim L\, \epsilon^{\alpha-1}  ~,~~~\alpha\ge 1 \quad \text{as}\quad \epsilon \to 0
\ee
BE condensation will occur.

We see that the existence of BE condensation { in the physics approach} depends solely on the {\it low-energy} behavior
of $\rho(\epsilon)$, while BE condensation according to CD depends solely on the {\it high-energy} behavior
of $\rho(\epsilon)$, a striking divergence of views. (The fact that the density $\rho(\epsilon)$ for many systems
is of a pure power low, and thus the two criteria converge, does not resolve the logical conundrum.)

This discrepancy also creates actual physical conflicts: for densities behaving as $\epsilon^{\alpha-1}$, $\alpha<1$ for
$\epsilon\to 0$ and as 
$\epsilon^{\alpha'-1}$, $\alpha'>1$ for $\epsilon\to \infty$, the physics approach predicts no BE condensation,
while the CD result assures there will be condensation. Conversely, for densities behaving as $\epsilon^{\alpha-1}$,
$\alpha>1$ for $\epsilon\to 0$ and as $\epsilon^{\alpha'-1}$, $\alpha'<1$ for $\epsilon\to \infty$, the physics approach
predicts condensation, while the CD result predicts no condensation (or, at least, is silent on the issue).

The above conflicts call for a resolution and, possibly, reconciliation. We shall provide these in the present note.
The upshot is that the physical implementation and interpretation of the (otherwise rigorous) mathematical results of CD
need to be properly done.

In a nutshell, the main issue is the following: in the CD analysis the asymptotic behavior as $\beta \to 0$ ($T \to \infty$)
is considered, which is assumed to be relevant for systems with $N \to \infty$ particles. In reality, neither $N$ nor
$T$ are infinite. Thus, the {\it finite} value of $\beta$ for which the limit $\beta \to 0$ is essentially attained, and the 
corresponding finite value of $N$, have to be compared to the actual number of particles and temperature. In this sense,
actual physical systems differ from mathematical limiting cases. As we will make clear, the low-energy behavior of $\ro$
will set the scale at which the mathematical limiting cases hold and will end up determining the system's condensation.

\section{Low-energy vs.\ high-energy spectral behavior}

The question of the existence of condensation  has to be properly posed. {\it Any} system can and will
condense if the temperature is small enough, that is, small compared to the gap above the ground state:
$k_{_B} T< \epsilon_1 - \epsilon_0$. True, physically relevant condensation should happen at temperatures macroscopically
larger than $\epsilon_1 - \epsilon_0$.
A related concept is that of extensivity: we expect the critical temperature  to be intensive, that is,
to depend on the density of particles $N/V$, with $V$ the system's volume (in a given dimension). This
concept is appropriately modified in the presence of external potentials, where other macroscopic parameters play the role of ``volume."

We will consider the grand canonical ensemble of non-interacting bosons at maximal chemical potential and will
assume that it reproduces
the distribution of particles as in the canonical ensemble, with the exception of the ground state, which will be treated
as a particle reservoir (this is a justified assumption, as explained earlier).
Therefore, assuming a discrete energy spectrum $\epsilon_j$, $j=0,1,2,\dots$ (and $\epsilon_0
=\mu = 0$), the maximal number of particles in levels other than the ground state  is
\be
N_\text{max} = \sum_{j=1}^\infty {1\over e^{\beta \e_j -1}}
\label{dis}\ee
Note that this sum is {\it always} finite, irrespective of the properties of the spectrum (again, strings excluded). It will,
therefore, imply the existence of a BE condensate at critical temperature $T_c = (k_B \beta_c)^{-1}$, at which $N_\text{max}$ equals the
actual number of particles  $N$, that is, $N_\text{max}(\beta_c) = N$. The issue is the magnitude of
this temperature: whether it is macroscopically larger than $\epsilon_1 - \epsilon_0$, and how it relates to the critical
temperature proposed by CD, which derives solely from the high-energy properties of the spectrum.

An explicit realization of $\epsilon_j$ that gives rise to a density $\ro \sim L\, \epsilon^{\alpha-1}$,
although by no means a unique one, is
\be
\epsilon_j = \left({\alpha j \over L}\right)^{1/\alpha}
\label{ej}\ee
This form will arise in our explicit calculations, in various adaptations.
The crucial point will be whether the sum in (\ref{dis}) can be approximated with an integral, as in (\ref{Nmax}). This is
entirely determined by the {\it low-energy} behavior of the spectrum.

In lieu of further elaboration, we will work out
explicitly two cases in which the physics approach and the CD results are in tension, that is, in which the low- and high-energy
behavior of the density $\ro$ come with different exponents, one bigger and one smaller than $1$.
For simplicity, we set
$k_{_B} = m =1$ in the sequel, with $m$ the mass of the bosons.

\subsection{Good low-energy, bad high-energy behavior}

Consider  a density of states that obeys
\bea
\ro &\sim& L\, \e^{\alpha -1} ~,\quad \alpha>1 \quad \text{as} \quad \e \to 0 \cr
&\sim& L' \e^{\alpha' -1}, \quad \alpha'\bb<\bb1 \quad \text{as} \quad \e \to \infty
\eea
A paradigm of such a system is a one-dimensional potential of the form
\bea
V(x) &=& a |x| ~, \quad~ |x|<{\ell/ 2} \cr
&=& \infty ~,\quad\quad |x|>{\ell/ 2}
\eea
representing a linear potential in a one-dimensional rigid box of size $\ell$.
For low energies the potential is confining, leading to BE condensation, while at higher energies it becomes
a box of size $\ell$, which does not condense. A WKB calculation yields
\vfill
\eject
\bea
\ro &=&{1\over h} \int dx\, dp\, \delta\bb\left(\epsilon - \half p^2 - V(x)\right) = {4\over h} \int_0^{x_o}\bb
 {dx \over \sqrt{2(\epsilon -V(x))}} \,, \quad V(x_o)=\epsilon \cr
 &=& {4\sqrt{2\epsilon}\over h a} ~,\quad\quad \epsilon < {a\ell\over 2}  \cr
 &=& {4\sqrt{2}\over ha} \big(\sqrt{\epsilon} - \sqrt{\epsilon - a\ell/2}\big) ~,\quad \epsilon >  {a\ell\over 2} 
\label{r11}\eea
Thus, \vskip -1cm
\bea
&&\epsilon \to 0 \, : ~\ro \sim {4\sqrt2 \over ha} \epsilon^{1/2} ~~~~~\text{so}~~ \alpha = {3\over 2}>1\cr
&&\epsilon \to \infty : ~\ro \sim {\ell\sqrt2 \over h} \epsilon^{-1/2} ~~\text{so}~~ \alpha' = {1\over 2} <1
\label{r1}\eea
This is the situation in which the physics approach  leads to condensation but the CD analysis does not.
The same WKB calculation yields for energy levels less than $a\ell/2$
\be
\epsilon_j = \left({3ha\over8\sqrt2}j\right)^{2/3}
\ee
while for $\epsilon > a\ell/2$, 
$\epsilon_j$ is the solution of a quartic equation, becoming at large energies
\be
\e_j^{3/2} - (\e_j-a\ell/2)^{3/2} = {3haj\over 8\sqrt 2} \quad\Rightarrow\quad 
\epsilon_j = \left({h j \over 2\sqrt2 \ell}\right)^2 \ \text{for}\ \e_j\gg a\ell
\ee
The maximal number of particles in levels other than the ground state is
\be
N_\text{max} = \sum_{j=1}^\infty {1\over e^{\beta \epsilon_j} -1} = \int_0^\infty {\ro \, d\epsilon \over e^{\beta \epsilon} -1}
\ee
The second equality holds for $\beta \ll (ha)^{-2/3}$, that is, $T$ much bigger than the gap above the ground
state, and relies on the fact that the sum near zero behaves as
\be
\sum_{j=1}^\infty {1\over e^{\beta \epsilon_j} -1} \sim \sum_j {1\over \epsilon_j} \sim \sum_j j^{-2/3}
\ee
Since the last expression diverges, the sum receives contributions from a large number of terms before the exponential
suppression kicks in and is thus well approximated by the corresponding integral.
Using the expression (\ref{r11}) for $\ro$ we obtain
\bea
N_\text{max} &=& {2\sqrt{2\pi}\over ha} T^{3/2} \Big( \zeta\big({\textstyle{3\over 2}}\big) - 
Li_{3\over 2} \big(e^{-a \ell/2T}\big)\Big) \cr
&=& {2\sqrt{2\pi}\over ha} \zeta\big({\textstyle{3\over 2}}\big) \, T^{3/2}\, ,\quad T \ll a\ell \cr
&=& {4\pi\over h} \sqrt{\ell \over a} \, T \, , \quad T \gg a\ell
\eea
We note that $N_\text{max}$ is finite for all temperatures, and thus predicts BE condensation. For small temperatures
we obtain the result for a linear confining potential. Importantly, for large temperatures we still obtain a finite
result. The critical temperature is
\bea
T_c &=& \left({N h a\over 2\sqrt{2\pi} \zeta\big({\textstyle{3\over 2}}\big)}\right)^{2/3} , \quad 1\ll N \ll h^{-1} a^{1/2} \ell^{3/2} \cr
&=& {N h\over 4\pi}\sqrt{a\over \ell} \ , \quad\quad N \gg h^{-1} a^{1/2} \ell^{3/2} 
\eea
and is much greater than the microscopic temperature $\epsilon_1 \bb-\bb \epsilon_0 \sim (ha)^{2/3}$ if
$a \gg h^2/\ell^3$, which is the case for a macroscopic parameter $a$ as realized in laboratory conditions.

\subsection{Bad low-energy, good high-energy behavior}

Consider, now, a system with a density of states that obeys
\bea
\ro &\sim& L\, \e^{\alpha -1} ~,\quad \alpha<1 \quad\quad \text{as} \quad \e \to 0 \cr
&\sim& L' \e^{\alpha' -1}, \quad \alpha'\bb>\bb1 \quad\quad \text{as} \quad \e \to \infty
\eea
A paradigm of such a system is a one-dimensional potential of the form
\bea
V(x) &=& 0 ~, \quad\quad\quad\quad |x|<{\ell/ 2} \cr
&=& a\Big(|x| - {\ell\over 2}\Big),\quad |x|>{\ell/ 2}
\eea
representing a linear confining potential with a flat bottom of size $\ell$.
For low energies the potential is flat, leading to no condensation, while at higher energies it becomes
linearly confining, leading to BE condensation. A WKB calculation yields
\bea
\ro &=&{1\over h} \int dx\, dp\, \delta\bb\left(\epsilon - \half p^2 - V(x)\right) = {4\over h} \int_0^{x_o}\bb
 {dx \over \sqrt{2(\epsilon -V(x))}} \,, \quad V(x_o)=\epsilon \cr
 &=& {\ell\sqrt 2\over h} \epsilon^{-1/2} + {4\sqrt 2\over h a} \epsilon^{1/2}
\label{rr}\eea
leading to low-energy behavior with $\alpha = {1\over 2} <1$ and high-energy behavior with $\alpha' = {3\over 2}>1$.
In this setting, the physics approach leads to no condensation while the CD analysis predicts
condensation. This is a subtler and more interesting situation.

A similar WKB calculation for the energy levels yields
\bea
&&\ell \epsilon_j^{1/2} +{4\over 3a} \epsilon_j^{3/2} = {h j \over 2\sqrt 2} ~, \quad \text{so}\cr
\epsilon_j &=& \left({h \over 2\sqrt2 \ell}j\right)^2~,\quad j \ll {\sqrt{a}\over h} \ell^{3/2}\cr
&=& \left({3ha\over 8\sqrt 2} j\right)^{2/3}~, \quad j \gg {\sqrt{a}\over h} \ell^{3/2}
\label{ejj}\eea
The maximal number of particles in levels other than the ground state is
\be
N_\text{max} = \sum_{j=1}^\infty {1\over e^{\beta \epsilon_j} -1} \neq \int_0^\infty {\ro \, d\epsilon \over e^{\beta \epsilon} -1}
\ee
The sum now {\it cannot} be well approximated by the corresponding integral, even for temperatures much larger than
the gap above the ground state $\e_1 -\e_0 \sim h^2/\ell^2$. Indeed, assuming $\beta h^2/\ell^2 \ll 1$, the sum near zero behaves as
\be
\sum_{j=1}^\infty {1\over e^{\beta \epsilon_j} -1} \sim \sum_j {1\over \epsilon_j} \sim \sum_j j^{-2}
\ee
It is finite, and thus the sum receives contributions mostly from the first few terms. To properly account for this,
we rewrite $N_\text{max}$ as
\be
N_\text{max} = \sum_{j=1}^J {1 \over e^{\beta \epsilon_j} -1} + \sum_{j=J+1}^\infty  {1 \over e^{\beta \epsilon_j} -1} \, ,
\quad 1\ll J\ll {\sqrt{a}\over h} \ell^{3/2} \, , \quad \beta \epsilon_J \ll 1
\ee
The existence of such a $J$ is guaranteed from the conditions $T \gg \epsilon_1 - \epsilon_0$ and 
$a \ell \gg \epsilon_1 - \epsilon_0$,
that is, the gap over the ground state is microscopically small compared to the
parameters of the system.
Then in the first sum
we can use the expression for $j\ll \sqrt{a}\ell^{3/2}/h$ in (\ref{ejj}), and the last sum is well approximated by the
corresponding integral since its terms vary little from $j$ to $j+1$. 
The finite sum becomes (for $J\gg 1$)
\be
{8\ell^2\over \beta h^2}  \sum_{j=1}^J {1\over j^2} = {8\ell^2\over \beta h^2} \Big({\pi^2\over 6} - {1\over J} \Big)
\ee
and expressing $J$ in terms of the corresponding energy $\epsilon_{J}$ the full expression for $N_\text{max}$ becomes
\be
N_\text{max} = {4\pi^2 \ell^2\over 3\beta h^2} - {2\sqrt 2 \ell\over \beta h \sqrt{\epsilon_J}} 
+ \int_{\epsilon_{\bb{_J}}}^\infty {\ro \, d\epsilon \over e^{\beta\epsilon} -1}
\ee
In the above, $\beta \epsilon_J \ll 1$, but the middle term rectifies the singularity of the integral as $\epsilon_J \to 0$
and their sum reaches a finite value as $\beta \e_J \to 0$. This can be seen by expressing $1/\sqrt{\epsilon_J}$
as an integral and combining the two integrands, which now give a finite integral.
Using (\ref{rr}) and changing integration variable, we have
\bea
N_\text{max} &=& {4\pi^2 \ell^2\over 3\beta h^2} + {\ell\sqrt 2 \over \beta^{1/2} h}\int_0^\infty ds\, 
s^{-1/2}\Big({1\over e^s -1} -{1\over s}\Big) + {4\sqrt2 \over \beta^{3/2} ha} \int_0^\infty ds {s^{1/2} \over e^s -1}\cr
&=&  {4\pi^2 \ell^2\over 3 h^2}T + {c\sqrt 2 \ell\over h} T^{1/2} + {2\sqrt{2\pi}\over ha}\zeta\big({\textstyle{3\over 2}}\big)\, T^{3/2}
\eea
where $c\simeq -2.588$ is the value of the middle integral. Since $\ell^2 T /h^2 \gg 1$ (by the assumption
$T \gg \epsilon_1 - \epsilon_0$), the second term is always subleading to the first one and can be neglected. Overall,
\be
N_\text{max} =  {4\pi^2 \ell^2\over 3 h^2}T + {2\sqrt{2\pi}\over ha}\zeta\big({\textstyle{3\over 2}}\big)\, T^{3/2}
\label{Nm}\ee
The first term is, under normal conditions, the dominant one. It is extremely large, and it is counted as ``infinity"
in the physics calculation in terms of the density of states. It predicts condensation at the extremely small
critical temperature
\be
T_c = {3h^2 N\over 4\pi^2 \ell^2}
\ee
The second term in (\ref{Nm}) is the result of CD. In order for it to dominate the previous term, we need
\be
T \gg {a^2 \ell^4 \over h^2}
\label{tan}\ee
This is an exceedingly {\it large} temperature, and it being the critical temperature of a BE condensate
would require
\be
N \gg {a^2 \ell^6 \over h^4}
\label{fan}\ee
which, under normal conditions, is a {\it fantastically} large number of particles.

As an example, assuming a small one-dimensional trap of $\ell = 1\, cm$ and potential slope $a$
of one electron Volt per micron trapping Helium-4 atoms,
the temperature (\ref{tan}) required for the CD result to
hold is larger than $10^{30} K$, more than a quadrillion times the temperature of the quark gluon plasma,
the hottest matter even created (for $10^{-22}$ seconds!), and the
number of particles (\ref{fan}) required is larger than $10^{42}$, or more than 10 quadrillion tons of mass!
By contrast, the low-energy part predicts a condensation for, say, $N=10^{6}$ particles (as in the sodium experiment)
at a critical temperature $T_c \sim 10^{-9} K$, which is a thousand times lower than the lowest temperature achievable
in a laboratory.

The moral of the story is that the result of CD, holding in the limit $\beta \to 0$,
requires $T$ to be way larger than any reasonable or achievable temperature, and taking it at face value
not only misses the fact that BE condensation in practice does not occur, but also that when it does occur experimentally it
happens at a critical temperature dramatically lower than the one implied by the CD analysis.

\section{Conclusions}

Bose-Einstein condensation still holds a few surprises and stretches our understanding of quantum statistical
mechanics in situations where the standard thermodynamic assumptions reach their limit, of which the reconciliation of
physics and mathematics results presented in the present work is an example. In the end, it is the low-energy
spectrum that determines physically relevant BE condensation and puts the results based on the high-energy spectrum to
their proper physical perspective.

The fact that BE condensation is ostensibly derived either from the low-energy or the high-energy behavior
of the spectrum is reminiscent of anomalies in quantum field theories, which can also be derived from high-energy
(Feynman diagrams) or low-energy (spectral flow) properties of the corresponding systems.
However, the present situation is qualitatively different: while the two approaches in field theory are related through topological
considerations and necessarily lead to the same results, they are genuinely distinct in the case of BE condensation,
as demonstrated by the two explicit examples
presented in this paper. No a priori topological or other principle relates the two ends of the spectrum, and the two
approaches can lead to physically different results, requiring the additional analysis presented in this paper.

The issue of condensation and the corresponding statistical fluctuations of particle number
become particularly interesting and challenging in the case of particles obeying exotic quantum statistics.
The case of particles obeying {\it inclusion} statistics, a form of enhanced Bose statistics, is especially interesting \cite{bibibis,bibi}.
Their statistical properties depend on a non-negative integer $g$, with $g=0$
corresponding to ordinary bosons. The probability distribution of occupation numbers $n_i$ on energy levels
$\e_i$ in the $n$-particle canonical ensemble is determined by the $n$-body partition function
 \be
 Z_g ={ \sum_{n_0, n_1,\ldots =0,\;\sum_{i=0} n_i=n
}^{\infty}{{\text {m}}_{g}(n_0,n_1,\ldots)}\,\rme^{-\beta\sum_{i=0} \epsilon_i n_i}}
\ee
which differs from the bosonic one by the presence of the extra multiplicity factors
\be\nonumber
{ {\text {m}}_{g} (n_0,n_1, \dots) =\;} { \prod_{i=0}^\infty} {[i,g+i] \over [i+1,g+i]}
\ee
where we defined the shorthand
\be
[i,j] := {{n_i + n_{i+1} + \cdots + n_j} \choose n_i \,,\, n_{i+1}\, ,\, \dots\, ,\, n_j} {~~\text{if}~~i<j, 
~~ [i,j] := 1 ~~\text{otherwise}}
\nonumber\ee
In the thermodynamic limit and in the absence of potentials, for any $g>0$ inclusion particles condense in one {\it less} space dimension
($d=2$ instead of $d=3$ for  BE condensation) and at higher critical temperatures
relative to  bosons. Further, a preliminary study of their microscopic properties
reveals that condensation does not only occur on the ground state, but rather a number of low-lying states share
the excess particles. The fluctuations of the number of particles on these states is also modified and requires
a nontrivial mathematical analysis. These and related questions are the subject of ongoing investigation.

\vskip 0.4cm

\noindent
{\bf Acknowledgements}

\noindent
The work of AP was supported in part by NSF under grant NSF-PHY-2112729 and by PSC-CUNY grants
65109-0053 and 6D136-0003. AP acknowledges the hospitality of the Laboratoire de Physique Th\'eorique et
Mod\`eles Statistiques of the Universit\'e Paris-Saclay where this research was initiated.

\noindent


\begin{thebibliography}{99}
\bibitem{Ketterle}  K. B. Davis, M. O. Mewes, M. R. Andrews, N. J. van Druten, D. S. Durfee, D. M. Kurn, and W. Ketterle, ``Bose-Einstein Condensation in a Gas of Sodium Atoms,'' Phys. Rev. Lett. 75, 3969 (1995).

\bibitem{Cornell}  M. H. Anderson,  J. R. Ensher, M. R. Matthews, C. E. Wieman,  E. A. Cornell, ``Observation of Bose-Einstein Condensation in a Dilute Atomic Vapor,'' Science 269 (5221): 198–201 (1995).

\bibitem{Diaconis}  
S. Chaterjee and P. Diaconis, ``Fluctuations of the Bose-Einstein condensate,'' J. Phys. A: Math. Theor. 47 085201 (2014).

\bibitem{Politzer}  H. D. Politzer, ``Condensate fluctuations of a trapped, ideal Bose gas,'' Phys.
Rev. A, 54 no. 6, 5048–5054  (1996).

\bibitem{bibibis}  S. Ouvry and  A. P. Polychronakos, ``Inclusion statistics and particle condensation in 2 dimensions,'' Phys. Rev. E 107, L062102 (2023).

\bibitem{bibi} S. Ouvry and  A. P. Polychronakos, ``Microscopic inclusion statistics in a discrete 1-body spectrum,'' Phys. Rev. E 110, 014140 (2024).

\end{thebibliography}
\end{document}